\documentclass[twocolumn,aps,prl,showpacs]{revtex4}
\usepackage[T1]{fontenc}
\usepackage[latin9]{inputenc}
\usepackage{amsmath}
\usepackage{amssymb}
\usepackage{graphicx}
\usepackage{esint}

\makeatletter
\@ifundefined{textcolor}{}
{%
 \definecolor{BLACK}{gray}{0}
 \definecolor{WHITE}{gray}{1}
 \definecolor{RED}{rgb}{1,0,0}
 \definecolor{GREEN}{rgb}{0,1,0}
 \definecolor{BLUE}{rgb}{0,0,1}
 \definecolor{CYAN}{cmyk}{1,0,0,0}
 \definecolor{MAGENTA}{cmyk}{0,1,0,0}
 \definecolor{YELLOW}{cmyk}{0,0,1,0}
}

\makeatother

\begin{document}

\title{Inhomogeneous Fulde-Ferrell superfluidity in spin-orbit coupled atomic
Fermi gases}

\author{Xia-Ji Liu$^{1}$}

\email{xiajiliu@swin.edu.au}

\author{Hui Hu$^{1}$}

\affiliation{$^{1}$ARC Centres of Excellence for Quantum-Atom Optics and Centre
for Atom Optics and Ultrafast Spectroscopy, Swinburne University of
Technology, Melbourne 3122, Australia}

\date{\today}
\begin{abstract}
Inhomogeneous superfluidity lies at the heart of many intriguing phenomena
in quantum physics. It is believed to play a central role in unconventional
organic or heavy-fermion superconductors, chiral quark matter, and
neutron star glitches. However, so far even the simplest form of inhomogeneous
superfluidity, the Fulde-Ferrell (FF) pairing state with a single
centre-of-mass momentum, is not conclusively observed due to the intrinsic
complexibility of any realistic Fermi systems in nature. Here we theoretically
predict that the controlled setting of ultracold fermionic atoms with
synthetic spin-orbit coupling induced by a two-photon Raman process,
demonstrated recently in cold-atom laboratories, provides a promising
route to realize the long-sought FF superfluidity. At experimentally
accessible low temperatures (i.e., $0.05T_{F}$, where $T_{F}$ is
the Fermi temperature), the FF superfluid state dominates the phase
diagram, in sharp contrast to the conventional case without spin-orbit
coupling. We show that the finite centre-of-mass momentum carried
by Cooper pairs is directly measurable via momentum-resolved radio-frequency
spectroscopy. Our work opens the way to direct observation and characterization
of inhomogeneous superfluidity.
\end{abstract}


\maketitle
Interacting Fermi systems with imbalanced spin populations are quite
ubiquitous in nature, with manifestations ranging from solid-state
superconductors to astrophysical objects \cite{Casalbuoni2004}. The
spin-imbalance disrupts the Bardeen-Cooper-Schrieffer (BCS) mechanism
of superconductivity, where fermions of opposite spin and momentum
form Cooper pairs. As a result, an exotic superconducting state, characterized
by Cooper pairs with a finite centre-of-mass momentum and spatially
non-uniform order parameter, may occur, as predicted by Fulde and
Ferrell nearly fifty years ago \cite{Fulde1964}. More complicated
inhomogeneous pairing states are also possible with the inclusion
of more and more momenta for the spatial structure of order parameter,
following the idea by Larkin and Ovchinnikov (LO) \cite{Larkin1964}.
These forms of inhomogeneous superfluidity, now referred to as FFLO
states, have attracted tremendous theoretical and experimental efforts
over the past five decades \cite{Radzihovsky2010}. Remarkably, to
date there is still no conclusive experimental evidence for FFLO superfluidity.
In solid state systems such as the organic superconductor $\lambda$-(BETS)$_{2}$FeCl$_{4}$
\cite{Uji2006} and the heavy fermion superconductor CeCoIn$_{5}$
\cite{Radovan2003,Kenzelmann2008}, the experimental difficulty arises
from unavoidable disorder effects and orbit/paramagnetic depairings
close to the upper critical Zeeman field.

An ultracold atomic Fermi gas has proven to be an ideal tabletop system
for the pursuit of FFLO superfluidity \cite{Radzihovsky2010}. Although
it is largely analogous to an electronic superconductor, the high
controllability in interactions, spin-populations and purity leads
to a number of unique experimental advances \cite{Bloch2008}. Indeed,
following theoretical proposals by Orso \cite{Orso2007} and the present
authors \cite{Hu2007,Liu2007,Liu2008}, strong experimental evidence
for the FFLO pairing has been observed in a one-dimensional spin-imbalanced
Fermi gas of $^{6}$Li atoms \cite{Liao2010}. In three dimensions
(3D), unfortunately, the FFLO phase occupies only an extremely small
volume in parameter space \cite{Sheehy2006,Hu2006} and thus is impossible
to observe experimentally \cite{Zwierlein2006,Partridge2006}.

\begin{figure}[t]
\begin{centering}
\includegraphics[clip,width=0.45\textwidth]{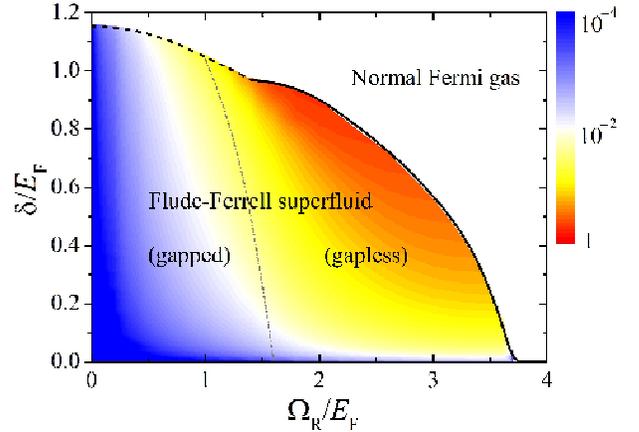} 
\par\end{centering}

\caption{(color online) Phase diagram of a 3D spin-orbit coupled atomic Fermi
gas at a broad Feshbach resonance and at the lowest experimentally
attainable temperature $0.05T_{F}$. The synthetic spin-orbit coupling
is induced by two counter-propagating Raman laser beams with strength
$\Omega_{R}$ and detuning $\delta$. Here we take the recoil momentum
$k_{R}=k_{F}$. A narrow LO wedge near $\Omega_{R}\sim0$ and $\delta\sim1.2E_{F}$
is not shown. By increasing $\delta$, the Fermi cloud changes from
a FF superfluid to a normal gas, via first-order (dashed line) and
second-order (solid line) transitions at low and high $\Omega_{R}$,
respectively. The FF superfluid can be either gapped or gapless. These
two phases are separated by the dot-dashed line. The color shows the
magnitude of the centre-of-mass momentum of Cooper pairs for the FF
superfluid, $q/k_{F}$. The BCS superfluid occurs at $\Omega_{R}=0$
or $\delta=0$ only.}

\label{fig1} 
\end{figure}

In this Letter, we predict that the FF superfluid could be easily
observed in a resonantly interacting 3D atomic Fermi gas with synthetic
spin-orbit coupling. This system was recently realized experimentally
at Shanxi University \cite{Wang2012} and at Massachusetts Institute
of Technology \cite{Cheuk2012}, using a two-photon Raman process.
Our main result is summarized in Fig. \ref{fig1}. At experimentally
accessible temperatures ($T\sim0.05T_{F}$) \cite{Partridge2006}
the FF state occupies a major part of the phase diagram. By tuning
the strength ($\Omega_{R}$) and detuning ($\delta$) of two Raman
laser beams, the centre-of-mass momentum of Cooper pairs $q$ can
be as large as the Fermi momentum $k_{F}$. We propose that such a
large FF momentum can be easily measured by momentum-resolved radio-frequency
spectroscopy (see Fig. \ref{fig4}) \cite{Stewart2008}. We note that
the emergence of a FF superfluid has also been predicted in a 3D atomic
Fermi gas in the presence of 2D Rashba spin-orbit coupling \cite{Zheng2012}
or 3D isotropic spin-orbit coupling \cite{Dong2013}, or in a two-dimensional
atomic Fermi gas with a two-photon Raman process \cite{Wu2012}. These
systems are yet to be realized experimentally. We also note that the
possibility of the FF superfluidity in our setting has been discussed
very recently by Vijay Shenoy \cite{Shenoy2012}.

We start by considering a 3D spin-orbit coupled spin-1/2 Fermi gas
of $^{6}$Li or $^{40}$K atoms near a broad Feshbach resonance \cite{Wang2012,Cheuk2012}.
Experimentally, the synthetic spin-orbit coupling is induced using
two counter-propagating Raman laser beams (i.e., along the $z$-direction)
that couple the two different spin states, following the same scenario
as in the NIST experiment for a $^{87}$Rb Bose-Einstein condensate
(BEC) \cite{Lin2011}. The Raman process can be described by, $(\Omega_{R}/2)\int d{\bf x[}\Psi_{\uparrow}^{\dagger}\left({\bf x}\right)e^{i2k_{R}z}\Psi_{\downarrow}\left({\bf x}\right)+$H.c.$]$,
where $\Psi_{\sigma}^{\dagger}\left({\bf x}\right)$ is the creation
field operator for atoms in the spin-state $\sigma$, $\Omega_{R}$
is the coupling strength of Raman beams, and $k_{R}$ $=2\pi/\lambda_{R}$
is the recoil momentum determined by the wave length $\lambda_{R}$
of the two beams. Thus, during the two-photon Raman process, a momentum
of $2\hbar k_{R}$ is imparted to an atom while its spin is changed
from $\left|\downarrow\right\rangle $ to $\left|\uparrow\right\rangle $.
This creates a correlation between the spin and orbital motion, which
can be seen more clearly by taking a gauge transformation, $\Psi_{\uparrow}({\bf x)}=e^{ik_{R}z}\psi_{\uparrow}({\bf x)}$
and $\Psi_{\downarrow}({\bf x)}=e^{-ik_{R}z}\psi_{\downarrow}({\bf x)}$.
Near a Feshbach resonance, the system may therefore be described by
a single-channel model Hamiltonian ${\cal H}=\int d{\bf x}[{\cal H}_{0}+{\cal H}_{int}]$,
where the single-particle part
\begin{equation}
{\cal H}_{0}=\left[\psi_{\uparrow}^{\dagger},\psi_{\downarrow}^{\dagger}\right]\left[\begin{array}{cc}
\hat{\xi}_{\mathbf{k}}+\lambda\hat{k}_{z}+\delta/2 & \Omega_{R}/2\\
\Omega_{R}/2 & \hat{\xi}_{\mathbf{k}}-\lambda\hat{k}_{z}-\delta/2
\end{array}\right]\left[\begin{array}{c}
\psi_{\uparrow}\\
\psi_{\downarrow}
\end{array}\right],\label{eq:spHami}
\end{equation}
and ${\cal H}_{int}=U_{0}\psi_{\uparrow}^{\dagger}\psi_{\downarrow}^{\dagger}\psi_{\downarrow}\psi_{\uparrow}\left({\bf x}\right)$
is the interaction Hamiltonian that describes the contact interaction
between two spin states with strength $U_{0}$. The interaction strength
should be expressed in terms of the \textit{s}-wave scattering length
$a_{s}$, i.e., $1/U_{0}=m/(4\pi\hbar^{2}a_{s})-1/V\sum_{{\bf k}}m/(\hbar^{2}k^{2})$,
which can be tuned precisely by sweeping an external magnetic field
around the Feshbach resonance \cite{Bloch2008}. Here $V$ is the
volume of the system. In the single-particle Hamiltonian (\ref{eq:spHami}),
$\hat{k}_{z}\equiv-i\partial_{z}$, $\hat{\xi}_{\mathbf{k}}\equiv-\hbar^{2}\nabla^{2}/(2m)-\mu$
after dropping a constant recoil energy, and $\delta$ is the two-photon
detuning from the Raman resonance. For convenience, we have defined
a spin-orbit coupling constant $\lambda\equiv\hbar^{2}k_{R}/m$. In
the Shanxi experiment with $^{40}$K atoms \cite{Wang2012}, the Fermi
wavelength is about $k_{F}\simeq1.6k_{R}$ and the coupling strength
$\Omega_{R}\simeq1.5E_{R}\simeq0.6E_{F}$, in units of the recoil
energy $E_{R}\equiv\hbar^{2}k_{R}^{2}/(2m)$ or Fermi energy $E_{F}\equiv\hbar^{2}k_{F}^{2}/(2m)$.

For the model Hamiltonian (\ref{eq:spHami}), most previous theoretical
studies focused on the case with $\delta=0$ and considered the standard
BCS superfluidity \cite{Iskin2011,Seo2012}. In this work, we are
interested in the FF pairing state with a single valued center-of-mass
momentum. Our investigation is motived by the interesting finding
that the two-body bound state at $\delta\neq0$ always acquires a
finite center-of-mass momentum along the $z$-axis \cite{Dong2012}.
At the many-body level, we therefore anticipate that at finite detunings
a FF superfluid could be more favorable than a BCS superfluid \cite{Shenoy2012}.

By assuming a FF-like order parameter $\Delta(\mathbf{x})=-U_{0}\left\langle \psi_{\downarrow}(\mathbf{x})\psi_{\uparrow}(\mathbf{x})\right\rangle =\Delta e^{iqz}$,
we consider the mean-field decoupling of the interaction Hamiltonian,
${\cal H}_{int}\simeq-[\Delta(\mathbf{x})\psi_{\uparrow}^{\dagger}(\mathbf{x})\psi_{\downarrow}^{\dagger}(\mathbf{x})+\textrm{H.c.}]-\Delta^{2}/U_{0}$.
Then, using a Nambu spinor $\Phi(\mathbf{x})\equiv[\psi_{\uparrow},\psi_{\downarrow},\psi_{\uparrow}^{\dagger},\psi_{\downarrow}^{\dagger}]$$^{T}$,
the total Hamiltonian can be written in a compact form, $\mathcal{H}=(1/2)\int d{\bf x}\Phi^{\dagger}(\mathbf{x})\mathcal{H}_{BdG}\Phi(\mathbf{x})-V\Delta^{2}/U_{0}+\sum_{\mathbf{k}}\hat{\xi}_{\mathbf{k}}$,
where the Bogoliubov Hamiltonian 
\begin{equation}
\mathcal{H}_{BdG}\equiv\left[\begin{array}{cccc}
\mathcal{S}_{\mathbf{k}}^{+} & \Omega_{R}/2 & 0 & -\Delta\left(\mathbf{x}\right)\\
\Omega_{R}/2 & \mathcal{S}_{\mathbf{k}}^{-} & \Delta\left(\mathbf{x}\right) & 0\\
0 & \Delta^{*}\left(\mathbf{x}\right) & -\mathcal{S}_{\mathbf{-k}}^{+} & -\Omega_{R}/2\\
-\Delta^{*}\left(\mathbf{x}\right) & 0 & -\Omega_{R}/2 & -\mathcal{S}_{-\mathbf{k}}^{-}
\end{array}\right]\label{eq:BdGHami}
\end{equation}
and $\mathcal{S}_{\mathbf{k}}^{\pm}\equiv\hat{\xi}_{\mathbf{k}}\pm\lambda\hat{k}_{z}\pm\delta/2$.
It is straightforward to diagonalize the Bogoliubov Hamiltonian $\mathcal{H}_{BdG}\Phi_{\mathbf{k\eta}}(\mathbf{x})=E_{\mathbf{k\eta}}\Phi_{\mathbf{k\eta}}(\mathbf{x})$
with quasiparticle wave-function $\Phi_{\mathbf{k\eta}}(\mathbf{x})=1/\sqrt{V}e^{i\mathbf{k\cdot}\mathbf{x}}[u_{\mathbf{k\eta\uparrow}}e^{+iqz/2},u_{\mathbf{k\eta\downarrow}}e^{+iqz/2},v_{\mathbf{k\eta\uparrow}}e^{-iqz/2},v_{\mathbf{k}\eta\downarrow}e^{-iqz/2}]^{T}$
and quasiparticle energy $E_{\mathbf{k}\eta}$ ($\eta=1,2,3,4$).
The mean-field thermodynamic potential $\Omega$ at temperature $T$
is then given by
\begin{eqnarray}
\frac{\Omega}{V} & = & \frac{1}{2V}\left[\sum_{\mathbf{k}}\left(\xi_{\mathbf{k}+\mathbf{q}/2}+\xi_{\mathbf{k}-\mathbf{q}/2}\right)-\sum_{\mathbf{k\eta}}E_{\mathbf{k}\eta}\right]\nonumber \\
 &  & -\frac{k_{B}T}{V}\sum_{\mathbf{k\eta}}\ln\left(1+e^{-E_{\mathbf{k}\eta}/k_{B}T}\right)-\frac{\Delta^{2}}{U_{0}}.\label{eq:Omega}
\end{eqnarray}
Note that the summation over the quasiparticle energy must be restricted
to $E_{\mathbf{k}\eta}\geq0$ because of an inherent particle-hole
symmetry in the Nambu spinor representation \cite{ParticleHoleSymmetry}.
For a given set of parameters (i.e, the temperature $T$, interaction
strength $1/k_{F}a_{s}$ etc.), different mean-field phases can be
determined using the self-consistent stationary conditions: $\partial\Omega/\partial\Delta=0$,
$\partial\Omega/\partial q=0$, as well as the conservation of total
atom number, $N=-\partial\Omega/\partial\mu$. At finite temperatures,
the ground state has the lowest free energy $F=\Omega+\mu N$.

Without loss of generality, hereafter we consider the resonance case
with a divergent scattering length $1/k_{F}a_{s}=0$ and set $T=0.05T_{F}$,
which is the lowest temperature that has been reported for Fermi atoms
\cite{Partridge2006}. According to the typical number of atoms in
experiments \cite{Wang2012,Cheuk2012}, we shall take $k_{R}=k_{F}$,
corresponding to a dimensionless spin-orbit coupling constant $\lambda k_{F}/E_{F}=2$.
The Raman coupling strength $\Omega_{R}$ and detuning $\delta$ can
be easily tuned experimentally. Thus, we focus on the phase diagram
as functions of $\Omega_{R}$ and $\delta$.

\begin{figure}[t]
\begin{centering}
\includegraphics[clip,width=0.48\textwidth]{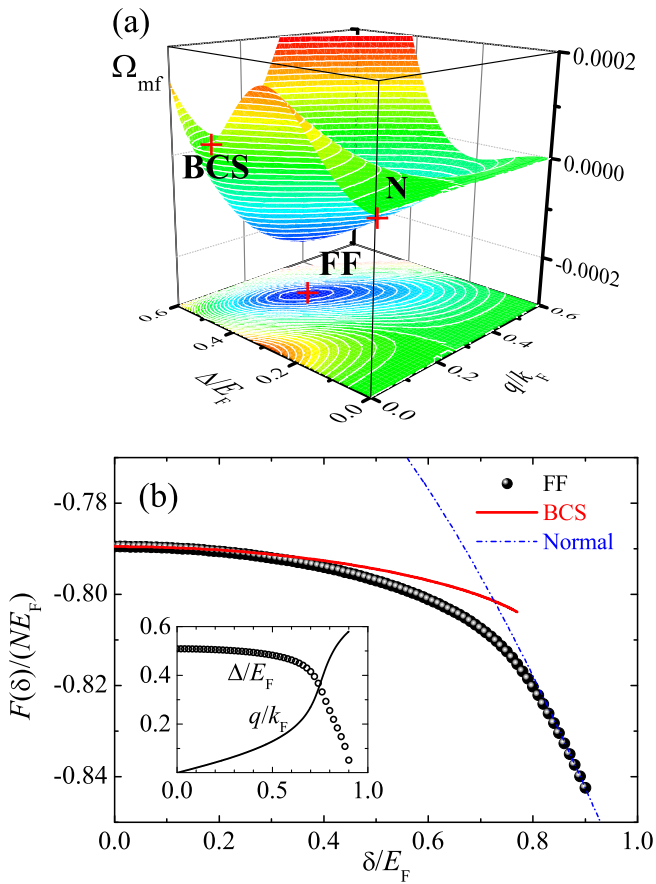} 
\par\end{centering}

\caption{(color online) (a) Landscape of the thermodynamic potential, $\Omega(\Delta,q)-\Omega(0,0)$,
at $\Omega_{R}=2E_{F}$ and $\delta=0.68E_{F}$. The chemical potential
is fixed to $\mu=-0.471E_{F}$. The competing ground states include
(i) a normal Fermi gas with $\Delta=0$; (ii) a fully paired BCS superfluid
with $\Delta\neq0$ and $q=0$; and (iii) a finite momentum paired
FF superfluid with $\Delta\neq0$ and $q\neq0$. (b) The free energy
of different competing states as a function of the detuning at $\Omega_{R}=2E_{F}$.
The inset shows the detuning dependence of the order parameter and
momentum of the FF superfluid state.}

\label{fig2}
\end{figure}

In general, for any set of parameters there are three competing ground
states that are stable against phase separation (i.e., $\partial^{2}\Omega/\partial\Delta^{2}\geq0$),
as shown in Fig. \ref{fig2}(a): normal gas ($\Delta=0$), BCS superfluid
($\Delta\neq0$ and $q=0$), and FF superfluid ($\Delta\neq0$ and
$q\neq0$). Remarkably, in the presence of spin-orbit coupling (i.e.,
$\Omega_{R}\neq0$) the FF superfluid is always more favorable in
energy than the standard BCS pairing state at finite detunings (Fig.
\ref{fig2}(b)). It is easy to check that the superfluid density of
the BCS pairing state in the axial direction becomes negative (i.e.,
$\partial\Omega/\partial q<0)$, signaling the instability towards
a FF superfluid. Therefore, experimentally the Fermi cloud would always
condense into a FF superfluid at finite Raman detunings. In Fig. \ref{fig1},
we report a low-temperature phase diagram that could be directly observed
in current experiments. The FF superfluid occupies the major part
of the phase diagram.

The greatly enlarged parameter space for a FF superfluid can be qualitatively
understood from the change of the Fermi surface due to spin-orbit
coupling, as discussed in the pioneering work by Barzykin and Gor'kov
\cite{Barzykin2002}. In our case, the single-particle energy spectrum
takes the form, $E_{\mathbf{k}\pm}=\hbar^{2}k^{2}/2m\pm\sqrt{(\Omega_{R}/2)^{2}+(\lambda k_{z}+\delta/2)^{2}}$,
where ``$\pm$'' stand for two helicity bands. At large spin-orbit
coupling ($\Omega_{R}\gg\lambda k_{F},\delta$), we find
\begin{equation}
E_{\mathbf{k}\pm}\simeq\frac{\hbar^{2}\left(k_{x}^{2}+k_{y}^{2}\right)}{2m}+\frac{\hbar^{2}}{2m}\left(k_{z}\pm\frac{q}{2}\right)\pm\frac{\Omega_{R}}{2}.
\end{equation}
where $q=2k_{R}\delta/\Omega_{R}$. Thus, the Fermi surfaces remain
approximately circular but the centers of the two Fermi surfaces are
shifted by $q/2$ along the $z$-axis in opposite directions. In this
situation, we may gain condensate energy by shifting the two Fermi
surfaces by an amount $\pm(q/2)\mathbf{e}_{z}$, respectively \cite{Agterberg}.
Here $\mathbf{e}_{z}$ is the unit vector in $z$-axis. The fermionic
pairing then occurs between single-particle states of $\mathbf{k}+(q/2)\mathbf{e}_{z}$
and $-\mathbf{k}+(q/2)\mathbf{e}_{z}$, leading to a FF order parameter
that has a spatial distribution $\Delta(\mathbf{x})=\Delta e^{iqz}$.
The direction of the FF momentum is uniquely fixed by the form of
spin-orbit coupling (i.e., $\lambda k_{z}$) and its magnitude is
roughly proportional to the detuning $\delta$.

At small spin-orbit coupling $\Omega_{R}\sim0$, however, the above
argument becomes invalid. Indeed, in the absence of spin-orbit coupling
($\Omega_{R}=0$) the instability to inhomogeneous superfluidity is
driven solely by the detuning. The FF superfluid occurs in a narrow
window (not shown in Fig. \ref{fig1}): $\delta_{cs}\leq\delta\leq1.204E_{F}$,
where $\delta_{cs}\simeq1.162E_{F}$ is the Clogston limit for a unitary
Fermi gas \cite{Hu2006}. The direction of the FF momentum cannot
be specified, indicating that a LO state with periodic stripe structure
is more preferable \cite{Burkhardt1994}. In our calculations, we
do not consider such a LO superfluid that exists in small wedge near
$\Omega_{R}\sim0$ and $\delta\sim1.2E_{F}$. Due to this simplification,
at low Raman coupling strengths the FF superfluid does not transform
continuously into a normal gas.

\begin{figure}[t]
\begin{centering}
\includegraphics[clip,width=0.48\textwidth]{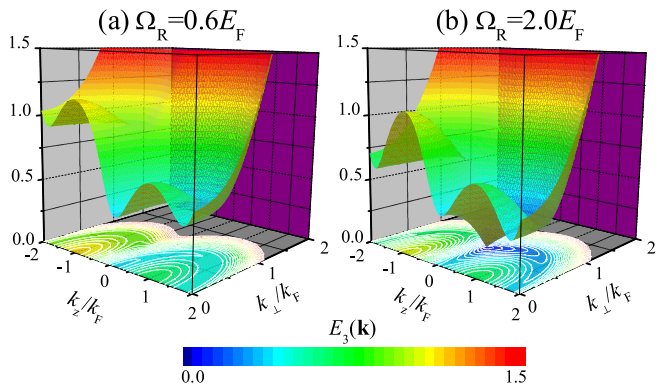} 
\par\end{centering}

\caption{(color online) The excitation spectrum $E_{3}(\mathbf{k})$ (in units
of $E_{F}$) as functions of $k_{\perp}=\sqrt{k_{x}^{2}+k_{y}^{2}}$
and $k_{z}$: (a) gapped spectrum at $\Omega_{R}=0.6E_{F}$ and (b)
gapless spectrum at $\Omega_{R}=2E_{F}$. Here we take $\delta=0.5E_{F}$.
Due to the particle-hole symmetry, $E_{2}(\mathbf{k})=-E_{3}(-\mathbf{k})$
and $E_{1}(\mathbf{k})=-E_{4}(-\mathbf{k})$. The gapped branches
$E_{1}(\mathbf{k})$ and $E_{4}(\mathbf{k})$ are not shown.}

\label{fig3} 
\end{figure}

The FF superfluid in Fig. \ref{fig1} can have different topological
structures in the quasiparticle excitation spectrum, as illustrated
in Fig. \ref{fig3}. At small spin-orbit couplings (Fig. \ref{fig3}(a)),
the FF superfluid has a negligible center-of-mass momentum and therefore
behaves very similar to a standard gapped BCS superfluid, while at
large spin-orbit coupling (Fig. \ref{fig3}(b)),  the sizable FF momentum
features a gapless spectrum. The nodal points with zero-excitation
energy are located at or close to the $k_{z}=0$ axis, along which
the dispersion relations take the form, $E(k_{\perp},0)=\pm\sqrt{(\lambda q+\delta)^{2}+\Omega_{R}^{2}}/2\pm\sqrt{\Delta^{2}+(\hbar^{2}k_{\perp}^{2}/2m+\hbar^{2}q^{2}/8m)^{2}}$.
Apparently, these nodal points would appear when the effective Zeeman
field $\sqrt{\delta^{2}+\Omega_{R}^{2}}/2$ is sufficiently large.
In Fig. \ref{fig1}, the gapped and gapless FF superfluids are separated
by a dot-dashed line. The topological transition from gapped to gapless
phases is continuous and could be revealed by thermodynamic measurements
through the atomic compressibility, spin susceptibility and momentum
distribution.

We now consider how to experimentally detect a FF superfluid through
momentum-resolved rf-spectroscopy \cite{Stewart2008}, which is a
cold-atom analog of the widely used angle-resolved photoemission spectroscopy
in solid-state systems. In such a measurement, one uses a rf field
with frequency $\omega$ to break a Cooper pair and transfer the spin-down
atom to an un-occupied third hyperfine state $\left|3\right\rangle $.
The momentum distribution of the transferred atoms is then measured
by absorption imaging after a time-of-flight expansion. The rf Hamiltonian
can be written as ${\cal H}_{rf}\propto\int d{\bf x[}e^{-ik_{R}z}\psi_{3}^{\dagger}\left({\bf x}\right)\Psi_{\downarrow}\left({\bf x}\right)+$H.c.$]$,
where $\psi_{3}^{\dagger}\left({\bf x}\right)$ is the field operator
which creates an atom in $\left|3\right\rangle $. The momentum transfer
$k_{R}{\bf e}_{z}$ in ${\cal H}_{rf}$ arises from the gauge transformation.
For a weak rf drive, the number of transferred atoms can be calculated
using linear response theory 
\begin{equation}
\Gamma\left(\mathbf{k},\omega\right)={\cal A}_{\downarrow\downarrow}\left[\mathbf{k}+(k_{R}-\frac{q}{2}){\bf e}_{z},\xi_{{\bf k}}-\omega\right]f\left(\xi_{{\bf k}}-\omega\right).\label{eq:rfstrength}
\end{equation}
Here ${\cal A}_{\downarrow\downarrow}(\mathbf{k},\omega)=\sum_{\eta}\left|u_{\mathbf{k}\eta\downarrow}\right|^{2}\delta(\omega-E_{\mathbf{k}\eta})$
is the single-particle spectral function of spin-down atoms, $\xi_{{\bf k}}\equiv\hbar^{2}k^{2}/(2m)-\mu$,
and$f(x)\equiv1/(e^{x/k_{B}T}+1)$ is the Fermi-Dirac distribution
function. For a FF superfluid, as Cooper pairs now carry a finite
center-of-mass momentum $\mathbf{q}=q\mathbf{e}_{z}$, the transferred
atoms acquire an overall momentum $\mathbf{q}/2$. As a result, there
is a $\mathbf{q}/2$ shift in the transferred strength Eq. (\ref{eq:rfstrength}),
which in principle could be experimentally measured.

\begin{figure}[t]
\begin{centering}
\includegraphics[clip,width=0.48\textwidth]{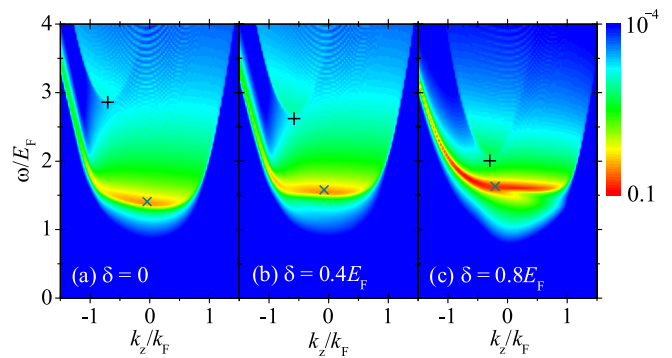} 
\par\end{centering}

\caption{(color online) Logarithmic contour plot of momentum-resolved rf spectroscopy:
number of transferred atoms $\Gamma(k_{z},\omega)$ at $\Omega_{R}=2E_{F}$
and at three detunings: (a) $\delta=0$ and $q=0$, (b) $\delta=0.4E_{F}$
and $q\simeq0.1k_{F}$, and (c) $\delta=0.8E$ and $q\simeq0.6k_{F}$.}

\label{fig4} 
\end{figure}

In Fig. 4, we report the momentum-resolved rf spectroscopy along the
$z$-direction $\Gamma(k_{z},\omega)\equiv\sum_{k_{\perp}}\Gamma({\bf k},\omega)$
at $\Omega_{R}=2E_{F}$ on a logarithmic scale. Quite generally, there
are two contributions to the spectroscopy, corresponding to two different
final states: after a Cooper pair is broken by a rf-photon, the remaining
spin-up atoms can be in different helicity bands \cite{Hu2012}. These
two contributions are well separated in the frequency domain, with
peak positions indicated by the symbols ``$+$'' and ``$\times$'',
respectively. Interestingly, at finite detuning with a sizable FF
momentum $\mathbf{q}$, the peak positions of the two contributions
are shifted roughly in opposite directions by an amount $\mathbf{q}/2$.
This provides clear evidence for observing the FF superfluid. 

In summary, we have shown that the condensate state of a spin-orbit
coupled Fermi gas is the long-sought inhomogeneous Fulde-Ferrell superfluid.
Our prediction can be readily examined in current experiments, where
a three-dimensional spin-orbit coupled atomic Fermi gas is created
using a two-photon Raman process \cite{Wang2012,Cheuk2012}. The finite
center-of-mass momentum carried by the inhomogeneous superfluid can
be unambiguously measured by momentum-resolved radio-frequency spectroscopy.
Our work complements relevant studies of solid-state systems, in which
inhomogeneous superfluidity was predicted to be enhanced by Rashba
spin-orbit interaction \cite{Barzykin2002}, but was difficult to
confirm experimentally. The controllable setting of a spin-orbit coupled
atomic Fermi gas opens a new direction to explore the fascinating
inhomogeneous superfluidity.

\textit{Acknowledgments}. We are grateful to Han Pu, Lin Dong, Peter
Hannaford for useful discussions. This research was supported by the
ARC Discovery Projects (DP0984637, DP0984522) and the NFRP-China 2011CB921502.


\begin{thebibliography}{10}
\bibitem{ParticleHoleSymmetry}It is straightforward to check that
for any particle state $[u_{\mathbf{k}\uparrow},u_{\mathbf{k}\downarrow},v_{\mathbf{k}\uparrow},v_{\mathbf{k}\downarrow}]$$^{T}$
with energy $E_{\mathbf{k}}\geq0$, there is a one-to-one corresponding
hole state $[v_{-\mathbf{k}\uparrow}^{*},v_{-\mathbf{k}\downarrow}^{*},u_{-\mathbf{k}\uparrow}^{*},u_{-\mathbf{k}\downarrow}^{*}]$$^{T}$
with energy $-E_{\mathbf{k}}$. These two states correspond to the
same physical solution.
\end{thebibliography}
\end{document}